\documentstyle[aps,eqsecnum,floats]{revtex}
\begin{document}
\twocolumn[\hsize\textwidth\columnwidth\hsize\csname 
           @twocolumnfalse\endcsname
\title{Regular coordinate systems for Schwarzschild and other
       spherical spacetimes}  
\author{Karl Martel and Eric Poisson}
\address{Department of Physics, University of Guelph, Guelph,
         Ontario, Canada N1G 2W1}
\maketitle
\begin{abstract}
The continuation of the Schwarzschild metric across the event horizon
is a well understood problem discussed in most textbooks on general
relativity. Among the most popular coordinate systems that are
regular at the horizon are the Kruskal-Szekeres and
Eddington-Finkelstein coordinates. Our first objective in this paper
is to popularize another set of coordinates, the Painlev\'e-Gullstrand
coordinates. These were first introduced in the 1920's, and have been
periodically rediscovered since; they are especially attractive and
pedagogically powerful. Our second objective is to provide
generalizations of these coordinates, first within the specific
context of Schwarzschild spacetime, and then in the context of more
general spherical spacetimes.    
\end{abstract}
\vskip 2pc]

\narrowtext

\section{Introduction} 

The difficulties of the Schwarzschild coordinates $(t,r,\theta,\phi)$
at the event horizon of a nonrotating black hole provide a vivid
illustration of the fact that in general relativity, the meaning of
the coordinates is not independent of the metric tensor
$g_{\alpha\beta}$. The Schwarzschild spacetime, whose metric is given 
by (we use geometrized units, so that $c = G = 1$) 
\begin{eqnarray}
ds^2 &=& -f\, dt^2 + f^{-1}\, dr^2 + r^2\, d\Omega^2, 
\nonumber \\ 
& & \label{1.1} \\
f &=& 1 - 2M/r \nonumber, 
\end{eqnarray}
where $d\Omega^2 = d\theta^2 + \sin^2\theta\, d\phi^2$, indeed gives
one of the simplest example of the failure of coordinates which have
an obvious interpretation in one region of the spacetime (the region
for which $r \gg 2M$), but not in another (the region for which
$r \leq 2M$). Understanding this failure of the ``standard'' 
coordinate system is one of the most interesting challenges in the
study of general relativity. Overcoming this obstacle is one of the
most rewarding experiences associated with learning the theory.  

Most textbooks on general relativity \cite{Schutz,MTW,Wald} discuss 
the continuation of the Schwarzschild solution across the event
horizon either via the Kruskal-Szekeres (KS) coordinates, or via the 
Eddington-Finkelstein (EF) coordinates; both coordinate systems 
produce a metric that is manifestly regular at $r=2M$. The main
purpose of this paper is to show that useful alternatives exist. One
of them, the Painlev\'e-Gullstand (PG) coordinates, are especially
simple and attractive, and we will consider them in detail. We will
also generalize them into a one-parameter family of coordinate
systems, and show that the EF and PG coordinates are members of this
family.    

In a pedagogical context, the KS coordinates come with several
drawbacks. First, the explicit construction of the KS coordinates is
relatively complicated, and must be carried out in a fairly long
series of steps. Second, the fact that $r$ is only implicitly defined
in terms of the KS coordinates makes working with them rather
difficult. Third, the manifold covered by the KS coordinates, with its
two copies of each surface $r = \mbox{constant}$, is unnecessarily
large for most practical applications; while the extension across the
event horizon is desirable, the presence of another asymptotic region
(for which $r \gg 2M$) often is not. While the KS coordinates are not
to be dismissed out of hand --- they do play an irreplaceable role in 
black-hole physics, and they should never be left out of a solid
education in general relativity --- we would advocate, for pedagogical
purposes and as a first approach to this topic, the construction of
simpler coordinate systems for extending the Schwarzschild spacetime
across the event horizon. 

A useful alternative are the EF coordinates $(v,r,\theta,\phi)$, in
which the metric takes the form
\begin{equation}
ds^2 = -f\, dv^2 + 2\, dvdr + r^2\, d\Omega^2. 
\label{1.2}          
\end{equation}
The new time coordinate $v$ is constant on ingoing, radial, null   
geodesics ($r$ decreases, $\theta$ and $\phi$ are constant); it is
related to the Schwarzschild time $t$ by $v = t + r^*$, where 
\begin{equation}  
r^* = \int \frac{dr}{f} = r + 2M \ln \biggl| \frac{r}{2M} - 1 \biggr|.   
\label{1.3}
\end{equation}
The metric of Eq.~(\ref{1.2}) is regular across the event
horizon. While its nondiagonal structure makes it slightly harder to
work with than the metric of Eq.~(\ref{1.1}), the fact that $r$
appears explicitly as one of the coordinates makes it much more
convenient than the KS version of the Schwarzschild metric. We believe
that in a pedagogical context, the Eddington-Finkelstein coordinates
should be introduced before the KS coordinates. 

Our first objective in this article is to popularize another set of
coordinates for Schwarzschild spacetime, and propose this system as a
useful alternative to the EF coordinates. These are the
Painlev\'e-Gullstrand (PG \cite{Painleve,Gullstrand}) coordinates
$(T,r,\theta,\phi)$. They are constructed and discussed in
Sec.~II. Our second objective is to provide generalizations of this 
coordinate system. In Sec.~III we discuss a one-parameter family of
PG-like coordinates for Schwarzschild spacetime. To the best of our
knowledge this family was first discovered by Kayll Lake 
in 1994 \cite{Lake}, but a related family of coordinates was
previously discussed by Gautreau and Hoffmann
\cite{GH,Gautreau,f0}. We show in Sec.~III that the PG and EF
coordinates are both members of Lake's family. In Sec.~IV we
generalize this family of coordinate systems to other spherical (and
static) spacetimes; equivalent coordinates were constructed, in a
two-dimensional context, by Corley and Jacobson \cite{f1}. In Sec.~V
we look back at our coordinates, and offer some additional comments
regarding their construction. In the Appendix we relate these
coordinate systems to the KS coordinates, and provide details
regarding the spacetime diagrams of Figs.~1 and 2.          

\section{Painlev\'e-Gullstrand coordinates} 

It is often a good strategy, when looking for regular coordinate
systems, to anchor the coordinates to a specific family of freely 
moving observers \cite{Novikov}. We shall employ this strategy
throughout this paper. The following derivation of the PG coordinates
can be found in the book by Robertson and Noonan
\cite{Robertson}. Other derivations can be found in
Refs.~\cite{GH,Gautreau,KW}, in which the PG coordinates were 
independently rediscovered.  

We consider observers which move along ingoing, radial, timelike
geodesics of the Schwarzschild spacetime ($r$ decreases, $\theta$ and
$\phi$ are constant). It is easy to check that in the standard
coordinates of Eq.~(\ref{1.1}), the geodesic equations can be
expressed in first-order form as 
\begin{equation}
\dot{t} = \frac{\tilde{E}}{f}, \qquad
\dot{r}^2 + f = \tilde{E}^2, 
\label{2.1}
\end{equation}
where an overdot denotes differentiation with respect to the
observer's proper time, and $\tilde{E} = E/m$ is the observer's
(conserved) energy per unit rest mass. (For a derivation, see Chap.~11
of Ref.~\cite{Schutz}, Chap.~25 of Ref.~\cite{MTW}, or Chap.~6 of
Ref.~\cite{Wald}.) We assume that $\dot{r} < 0$, and the energy
parameter is related to the observer's initial velocity $v_\infty$ ---
the velocity at $r = \infty$ --- by  
\begin{equation}
\tilde{E} = \frac{1}{\sqrt{1-{v_\infty}^2}}. 
\label{2.2}
\end{equation} 

In this section we specialize to the particular family of observers
characterized by $\tilde{E} = 1$; our observers are all starting at
infinity with a zero initial velocity: $v_\infty = 0$. For these
observers, the geodesic equations reduce to $\dot{t} = 1/f$ and
$\dot{r} = -\sqrt{1-f}$. We notice that $u_\alpha$, the covariant
components of the observer's four-velocity, whose contravariant
components are $u^\alpha = (\dot{t}, \dot{r}, 0, 0)$, is given by
$u_\alpha = (-1,-\sqrt{1-f}/f,0,0)$. This means that $u_\alpha$ is
equal to the gradient of some time function $T$:   
\begin{equation}
u_\alpha = -\partial_\alpha T,
\label{2.3}
\end{equation}
where 
\begin{equation}
T = t + \int \frac{ \sqrt{1-f} }{f}\, dr. 
\label{2.4}
\end{equation}
Integration of the second term is elementary, and we obtain
\begin{equation}
T = t + 4M \biggl( \sqrt{r/2M} + \frac{1}{2} \ln 
\biggl| \frac{ \sqrt{r/2M} - 1 }{ \sqrt{r/2M} + 1 } \biggr| \biggr).  
\label{2.5}
\end{equation}
This shall be our new time coordinate, and $(T,r,\theta,\phi)$ are
nothing but the PG coordinates. It should be clear that the key to the
construction of the PG coordinates is the fact that the four-velocity
can be expressed as in Eq.~(\ref{2.3}). In Sec.~V we will explain how
this equation comes about. 

Going back to Eq.~(\ref{2.4}), we see that $dt = dT -
f^{-1}\sqrt{2M/r}\, dr$. Substituting this into Eq.~(\ref{1.1}) gives  
\begin{equation}
ds^2 = -f\, dT^2 + 2 \sqrt{2M/r}\, dT dr + dr^2 + r^2\, d\Omega^2. 
\label{2.6}
\end{equation}
This is the Schwarzschild metric in the PG coordinates. An equivalent
way of expressing this is 
\begin{equation}
ds^2 = -dT^2 + \Bigl( dr + \sqrt{2M/r}\, dT \Bigr)^2 + r^2\,
d\Omega^2.
\label{2.7}
\end{equation}
This metric is manifestly regular at $r = 2M$, in correspondence with
the fact that our observers do not consider this surface to be in any
way special. (The metric is of course still singular at $r=0$.) While
the metric is now nondiagonal, it has a remarkably simple form. It is
much simpler than the Kruskal-Szekeres metric, and we believe that it
provides a useful alternative to the Eddington-Finkelstein form of the
metric, Eq.~{1.2}. 

In Fig.~1 we show several surfaces $T = \mbox{constant}$ in a Kruskal 
diagram. The construction is detailed in the Appendix. The diagram
makes it clear that the PG coordinates do not extend inside the past
horizon of the Schwarzschild spacetime --- the ``white-hole region''
is not covered. The reason for this is that our observers fall inward
from infinity and end up crossing the {\it future}, but not the 
{\it past}, horizon. By reversing the motion (choosing the opposite
sign for $\dot{r}$), we would obtain alternative coordinates that
extend within the past horizon but do not cover the black-hole region
of the spacetime. While the PG coordinates do not cover the entire KS
manifold, they do cover the two most interesting regions of the
maximally extended Schwarzschild spacetime.  

\begin{figure}
\vspace*{2.5in}
\special{hscale=33 vscale=33 hoffset=-13.0 voffset=190.0
         angle=-90.0 psfile=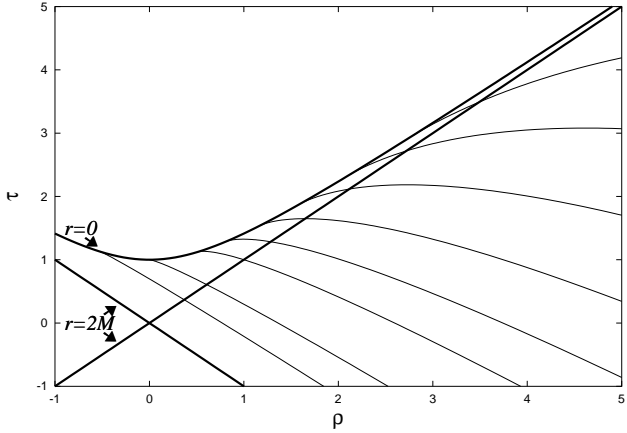}
\caption{Surfaces $T = \mbox{constant}$ in a Kruskal diagram. The
vertical and horizontal axes correspond to the Kruskal $\tau$ and
$\rho$ coordinates, respectively; these are defined in the
Appendix. The thick, diagonal lines represent the two copies of the
surface $r = 2M$; the future horizon is oriented at $+45$ degrees,
while the past horizon is oriented at $-45$ degrees. The thick,
hyperbolic line represents the curvature singularity at $r=0$. The
thin lines represent the surfaces $T = \mbox{constant}$. From the
bottom up we display the surfaces $T=-2M$, $T=0$, $T=2M$, $T=3M$,
$T=4M$, $T=5M$, $T=6M$, and $T=7M$.} 
\end{figure}

Equations (\ref{2.6}) and (\ref{2.7}) reveal the striking property
that the surfaces $T = \mbox{constant}$ are intrinsically flat:  
Setting $dT = 0$ returns $ds^2 = dr^2 + r^2\, d\Omega^2$, which is the 
metric of flat, three-dimensional space in spherical polar
coordinates. The information about the spacetime curvature is
therefore entirely encoded in the ``shift vector'', the off-diagonal  
component of the metric tensor. We consider this aspect of the PG
coordinates to be their most interesting property.      

We note that it is possible to construct PG-like coordinates for the
nonspherical Kerr spacetime. This was carried out by C.~Doran in a
recent paper \cite{D}. 

\section{Generalization to other observers} 

It is easy to generalize the preceding construction to other families 
of freely moving observers. In this section we consider families such
that $\tilde{E}$ is the same for all observers within the family, but
not equal to unity (as in Sec.~II). Each family is therefore
characterized by its unique value of the energy parameter. We find it
convenient to use instead the parameter $p$, related to the energy and 
initial-velocity parameters by
\begin{equation}
p = \frac{1}{\tilde{E}^2} = 1 - {v_\infty}^2. 
\label{3.1}
\end{equation}
We take $p$ to be in the interval $0 < p \leq 1$, with $p=1$ taking us 
back to the PG coordinates \cite{f2}. To each value of $p$ in this
interval corresponds a family of observers, and a distinct coordinate 
system. We are therefore constructing a one-parameter family of
PG-like coordinates for Schwarzschild spacetime. 

With the geodesic equations now given by $\dot{t} = 1/(\sqrt{p} f)$
and $\dot{r} = -\sqrt{1-pf}/\sqrt{p}$, we find that $u_\alpha$ is now
equal to a constant times the gradient of a time function $T$:
\begin{equation}
u_\alpha = - \frac{1}{\sqrt{p}}\, \partial_\alpha T,
\label{3.2}
\end{equation}
with 
\begin{equation}
T = t + \int \frac{\sqrt{1-pf}}{f}\, dr. 
\label{3.3}
\end{equation}
Integration of the second term doesn't present any essential
difficulties, and we obtain
\begin{eqnarray}
T &=& t + 2M \biggl( \frac{1-pf}{1-f} 
+ \ln \biggl| \frac{1 - \sqrt{1-pf}}{1 + \sqrt{1-pf}} \biggr| 
\nonumber \\ & & \mbox{}  
- \frac{1-p/2}{\sqrt{1-p}} 
\ln \biggl| \frac{\sqrt{1-pf}
-\sqrt{1-p}}{\sqrt{1-pf}+\sqrt{1-p}} \biggr| \biggr).  
\label{3.4}
\end{eqnarray} 
This shall be our new time coordinate. In Sec.~V we will return to the
question of the origin of Eq.~(\ref{3.2}). 

With $dt$ now equal to $dT - f^{-1}\sqrt{1-pf}\, dr$, we find that the
Schwarzschild metric takes the form  
\begin{equation}
ds^2 = -f\, dT^2 + 2\sqrt{1-pf}\, dTdr + p\, dr^2 + r^2\, d\Omega^2, 
\label{3.5}
\end{equation}
or
\begin{equation}
ds^2 = -\frac{1}{p}\, dT^2 + p \biggl( dr 
+ \frac{1}{p} \sqrt{1-pf}\, dT \biggr)^2 + r^2\, d\Omega^2.
\label{3.6}
\end{equation}
This metric is still regular at $r=2M$, although it is now slightly
more complicated than the PG form.  

In Fig.~2 we show several surfaces $T = \mbox{constant}$ in a Kruskal 
diagram, for several values of $p$. This construction is detailed in 
the Appendix. 

\begin{figure}
\vspace*{2.5in}
\special{hscale=33 vscale=33 hoffset=-13.0 voffset=190.0
         angle=-90.0 psfile=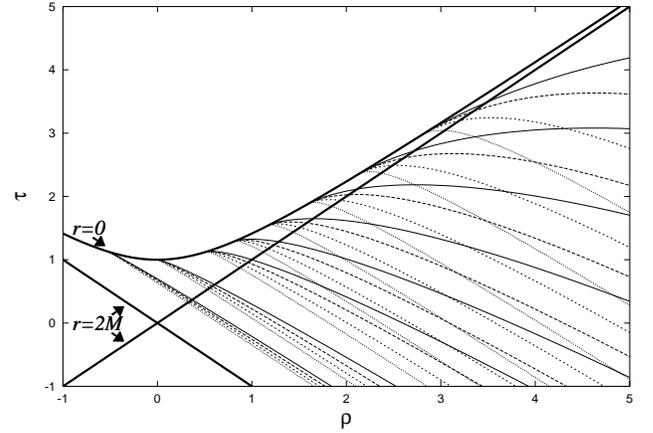}
\caption{Surfaces $T = \mbox{constant}$ in a Kruskal diagram. The axes
and the meaning of the thick lines are explained in the caption of
Fig.~1. The eight bundles of thin lines refer to the same values of
$T$ as in Fig.~1, from $T = -2M$ (bottom bundle) to $T = 7M$ (top
bundle). Within a single bundle, each of the four lines come with a 
distinct value of the parameter $p$. The solid line corresponds to
$p=1$, the long-dashed line corresponds to $p=3/4$, the short-dashed
line corresponds to $p=1/2$, and the dotted line corresponds to
$p=1/4$.}  
\end{figure}

In this generalization of the PG coordinates, the surfaces $T =
\mbox{constant}$ are no longer intrinsically flat. The induced metric 
is now $ds^2 = p\, dr^2 + r^2\, d\Omega^2$, and although the factor of  
$p$ in front of $dr^2$ looks innocuous, it is sufficient to produce a
curvature. It may indeed be checked that the Riemann tensor associated
with this metric is nonzero. The only nonvanising component is 
$R^\phi_{\ \theta\phi\theta} = -(1-p)/p$, and $R^{abcd} R_{abcd} =
4(1-p)^2/(pr^2)^2$.   

It is instructive to go back to Eq.~(\ref{3.4}) and check that in the
limit $p \to 1$, $T$ reduces to the expression of
Eq.~(\ref{2.5}). (This must be done as a limiting procedure, because
$T$ is ambiguous for $p = 1$.) Taking the limit gives
\begin{equation}
\lim_{p \to 1} T = t + 2M \biggl( \frac{2}{\sqrt{1-f}} 
+ \ln \biggl| \frac{1-\sqrt{1-f}}{1+\sqrt{1-f}} \biggr| \biggr), 
\label{3.7}
\end{equation}
which is indeed equivalent to Eq.~(\ref{2.5}). The PG coordinates are
therefore a member of our one-parameter family. 

Another interesting limit is $p \to 0$, which corresponds to
$\tilde{E} \to \infty$, or $v_\infty \to 1$. In this limit, our
observers start at infinity with a velocity nearly equal to the speed
of light. Starting from Eq.~(\ref{3.4}) we have
\begin{equation}
\lim_{p \to 0} T = t + 2M \biggl( \frac{1}{1-f} 
+ \ln \biggl| \frac{f}{1-f} \biggr| \biggr) = t + r^*, 
\label{3.8}     
\end{equation}
where we have compared with Eq.~(\ref{1.3}). Thus, $T = v$ in the
limit $p \to 0$, and our generalized PG coordinates reduce to the
Eddington-Finkelstein coordinates of Eq.~(\ref{1.2}). This is not
entirely surprising, in view of the fact that our observers become
light-like in this limit. The EF coordinates, therefore, are also a
(limiting) member of our one-parameter family.   

We have constructed an interpolating family of coordinate systems for
Schwarzschild spacetime; as the parameter $p$ varies from 1 to 0, the
coordinates go smoothly from the Painlev\'e-Gullstrand coordinates to
the Eddington-Finkelstein coordinates. This one-parameter family of
coordinate systems was first discovered by Kayll Lake \cite{Lake}, but
a related family of coordinates, corresponding to $p > 1$, were
previously introduced by Gautreau and Hoffmann
\cite{GH,Gautreau,f0}. Lake obtained the new coordinates
by solving the Einstein field equations for a vacuum, spherical
spacetime in a coordinate system involving $r$ and an arbitrary time
$T$. The intimate relation between his coordinates and our families of
freely moving observers remained unnoticed by him. 

\section{Generalization to other spacetimes}

The coordinates constructed in the previous two sections can be
generalized to other static and spherically symmetric spacetimes. In 
the usual Schwarzschild coordinates, we write the metric as 
\begin{equation}
ds^2 = -e^{2\psi} f\, dt^2 + f^{-1}\, dr^2 + r^2\, d\Omega^2, 
\label{4.1}
\end{equation}
where $f$ and $\psi$ are two arbitrary functions of $r$. Under the
stated symmetries, Eq.~(\ref{4.1}) gives the most general form for the
metric. We assume that the spacetime is asymptotically flat, so that
$f \to 1$ and $\psi \to 0$ as $r \to \infty$. If the spacetime
possesses a regular event horizon at $r = r_0$, then $f(r_0) = 0$
and $\psi$ must be nonsingular for all values of $r \neq 0$.      

The geodesic equations are now
\begin{equation} 
\dot{t}= \frac{\tilde{E}}{e^{2\psi} f}, \qquad
\dot{r}^2 + f = e^{-2\psi} \tilde{E}^2, 
\label{4.2}
\end{equation}
where $\tilde{E}$ is still the conserved energy per unit rest
mass. Re-introducting $p = 1/\tilde{E}^2$, we find that the covariant 
components of the four-velocity can be again expressed as in
Eq.~(\ref{3.2}), with a time function $T$ now given by 
\begin{equation}
T = t + \int \frac{\sqrt{e^{-2\psi} - p f}}{f}\, dr. 
\label{4.3}
\end{equation}
The second term can be integrated if $f$ and $\psi$ are
known. Rewriting the metric of Eq.~(\ref{4.1}) in terms of $dT$ yields  
\begin{eqnarray}
ds^2 &=& -f e^{2\psi}\, dT^2 
+ 2 e^{2\psi} \sqrt{e^{-2\psi} - p f}\, dT dr
\nonumber \\ & & \mbox{}
 + p e^{2\psi}\, dr^2 + r^2\, d\Omega^2, 
\label{4.4}
\end{eqnarray}
or
\begin{eqnarray}
ds^2 &=& -\frac{1}{p}\, dT^2 + p e^{2\psi} \biggl( dr 
+ \frac{1}{p}\, \sqrt{e^{-2\psi} - p f}\, dT \biggr)^2 
\nonumber \\ & & \mbox{} 
+ r^2\, d\Omega^2. 
\label{4.5}
\end{eqnarray}
This metric is manifestly regular at an eventual event horizon, at
which $f$ vanishes.   

The surfaces $T = \mbox{constant}$ have an induced metric given by
$ds^2 = p e^{2\psi}\, dr^2 + r^2\, d\Omega^2$. Unless $\psi = 0$ and  
$p = 1$, these surfaces are not intrinsically flat \cite{f3}.  

\section{Final comments} 

In all the cases considered in Secs.~II, III, and IV, the construction
of our coordinate systems relied on the key fact that the 
four-velocity could be expressed as $u_\alpha =
-\partial_\alpha T/\sqrt{p}$, with $p$ a constant. [This is
Eq.~(\ref{3.2}), and in Sec.~II, $p$ was set equal to unity.] This
property is remarkable, and it seems to follow quite accidentally from
the equations of motion. There is of course no accident, but the point
remains that not every four-velocity vector can be expressed in this
form.  

A standard theorem of differential geometry (for example, see Appendix
B of Ref.~\cite{Wald}) states that for $u_\alpha$ to admit the form of
Eq.~(\ref{3.2}), it must satisfy the equations $u^\alpha_{\ ;\beta}
u^\beta = 0$ and $u_{[\alpha;\beta} u_{\gamma]} = 0$, in which a
semicolon denotes covariant differentiation and the square brackets
indicate complete antisymmetrization of the indices. The second
equation states that the world lines are everywhere orthogonal to a
family of spacelike hypersurfaces, the surfaces of constant $T$. This
ensures that the four-velocity can be expressed as $u_\alpha = -\mu\, 
\partial_\alpha T$, for some {\it function}  $\mu(x^\alpha)$. In
general, this function is not a constant, and we do yet have
Eq.~(\ref{3.2}). For this we need to impose also the first equation,
which states that the motion is geodesic. When both equations hold we
find that $\mu = \mbox{constant}$, and this gives us Eq.~(\ref{3.2}).   

In our constructions, we have enforced the geodesic equation by
selecting freely moving observers. By selecting {\it radial}
observers, we have also enforced the condition that the geodesics be
hypersurface orthogonal. Our strategy for constructing coordinate
systems is therefore limited to radial, freely moving observers in
static, spherically-symmetric spacetimes; it may not work for more
general motions and/or more general spacetimes.  

\section*{Acknowledgments}

This work was supported by the Natural Sciences and Engineering
Research Council of Canada. We are grateful to Kayll Lake, Ted
Jacobson, and an anonymous referee for discussions and comments on the
manuscript.  

\appendix
\section{Kruskal diagrams} 

The Kruskal diagrams of Figs.~1 and 2 are constructed as follows. 

From the Schwarzschild coordinates $t$ and $r$ we define two null  
coordinates, $u = t - r^*$ and $v = t + r^*$, where $r^*$ is given by 
Eq.~(\ref{1.3}). From these we form the null KS coordinates, $V =
e^{v/4M}$ and $U = \mp e^{-u/4M}$, where the upper sign refers to the
region $r > 2M$ of the Schwarzschild spacetime, while the lower sign
refers to $r < 2M$. From this we derive the relations
\begin{equation}
UV = -e^{r/2M} \biggl( \frac{r}{2M} -1 \biggr)
\label{A.2}
\end{equation}
and 
\begin{equation}
\frac{V}{U} = \mp e^{t/2M}. 
\label{A.3}
\end{equation}
Timelike and spacelike KS coordinates are then defined by $V =
\tau + \rho$ and $U = \tau - \rho$. In our spacetime
diagrams, the $\tau$ axis runs vertically, while the $\rho$ axis
runs horizontally. The future horizon is located at $U = 0$, and the
past horizon is at $V = 0$. The curvature singularity is located at
$UV = 1$.  

We express the time function of Eq.~(\ref{3.4}) as
\begin{equation}
T = t + r^* + S(r),
\label{A.4}
\end{equation}
where $S(r)$ is the function of $r$ that results when the second term
of Eq.~(\ref{3.4}) is shifted by $-r^*$, as given in
Eq.~(\ref{1.3}); this function is regular at $r=2M$. With this
definition we have $v = T-S$, $u = T-S-2r^*$, as well as   
\begin{equation}
V = e^{T/4M} e^{-S/4M}
\label{A.5}
\end{equation}
and
\begin{equation}
U = - e^{r/2M} (r/2M - 1) e^{-T/4M} e^{S/4M}. 
\label{A.6}
\end{equation}
The surfaces $T = \mbox{constant}$ give rise to parametric equations
of the form $V(r)$ and $U(r)$, which are obtained from
Eqs.~(\ref{A.5}) and (\ref{A.6}) by explicitly evaluating the function
$S(r)$. In these equations, $r$ can be varied from zero to an
arbitrarily large value without difficulty. The diagrams of Figs.~1
and 2 are then produced by switching to the coordinates $\bar{t}$ and
$\bar{r}$ and plotting the parametric curves.


\begin{references}
\bibitem{f0} 
   The Gautreau-Hoffmann coordinates are very similar to the
   coordinates of Sec.~III. They also constitute a one-parameter
   family of coordinate systems, and their parameter $R_i$ is related
   to our $p$ by $R_i = 2Mp/(p-1)$. The difference lies with the fact
   that while their $R_i$ is not meant to be negative, our $p$ is
   restricted to the interval $0 < p \leq 1$. These are mutually
   exclusive statements. But the point remains that formally, the
   coordinates of Sec.~III are identical to the Gautreau-Hoffmann
   coordinates for $R_i < 0$. Note that for $R_i > 0$ (the case
   considered by Gautreau and Hoffmann in Ref.~[7]), the surfaces of
   constant time extend only up to $r=R_i$; they do not reach
   infinity.
\bibitem{f2} 
   For $p > 1$, or $\tilde{E} < 1$, the motion does not extend to
   infinity but starts at a turning-point radius $r_{\rm max}$ defined
   by $f(r_{\rm max}) = \tilde{E}^2$. We exclude these cases because
   the resulting surfaces $T = \mbox{constant}$ would not extend
   beyond $r = r_{\rm max}$, and the new coordinates would not be
   defined everywhere. These cases, however, are discussed in
   Refs.~[7,8] above. See the remark in Ref.~[9].
\bibitem{f3}
   We exclude the case $\psi = \mbox{constant} \neq 0$, because the
   factor $e^{2\psi}$ in front of $dt^2$ in Eq.~(\ref{4.1}) can then
   be absorbed into a rescaling of the time coordinate. In other
   words, there is no loss of generality involved in setting $\psi =
   0$ if $\psi$ is initially an arbitrary constant.
\end{references}
\end{document}